\documentclass[prl]{revtex4-2}
\usepackage{graphicx}
\usepackage{commath}
\begin{document}
\title{Bounded System of Monopole and Half-monopole in the Weinberg-Salam Model}
\author{Dan Zhu, Khai-Ming Wong and Timothy Tie}
\affiliation{School of Physics, Universiti Sains Malaysia, 11800 USM, Penang}

\begin{abstract}
\textbf{Abstract}. In this work, we study the one plus half-monopole configuration in Weinberg-Salam model, covering $\phi$-winding number $n$ = 1. We observed that while the finite separation between the one-monopole and the half-monopole becomes larger as compared to the same configuration in SU(2) Yang-Mills-Higgs theory, a flux tube is established, creating a bound-state of one-monopole and half-monopole in Weinberg-Salam model. There is no electromagnetic current loop circulating the pair of one-monopole and half-monopole, but the system possesses non-vanishing magnetic dipole moment. The non-Abelian gauge potential and the electromagnetic gauge potential are singular along the negative $z$-axis, but the total energy is finite. The solutions are investigated by fixing the Weinberg angle while varying the Higgs self-coupling constant and vice versa. It is shown that this configuration in Weinberg-Salam model possesses different behaviours than its counterpart in SU(2) Yang-Mills-Higgs theory.
\end{abstract}

\maketitle

\section{Introduction}
The idea of magnetic monopole was first introduced into the Maxwell's theory by P.A.M. Dirac \cite{PaulDirac} in 1931. It is a point magnetic charge with a semi-infinite string attached to it and it possesses infinite energy. An idea to restore the symmetry between electricity and magnetism which the current Maxwell's equations are seemingly missing. It possesses a magnetic charge of $\frac{2\pi n}{e}$, where $e$ is the unit electric charge and $n$ is an integer. The fact that electric charges are quantized naturally and there exists no other explanation for this quantization makes magnetic monopole a very important particle that has yet to be discovered. \par

In 1969, the first non-Abelian magnetic monopole was found by Wu and Yang \cite{WuYang} in the pure SU(2) Yang-Mills theory with only a single point singularity present in the solution. However, just like the Dirac monopole, the energy of the Wu-Yang monopole is still infinite due to the presence of singularities. It was not until 1974 that the first finite energy magnetic monopole was finally found by 't Hooft and Polyakov \cite{1} independently in the SU(2) Yang-Mills-Higgs theory. The mass of the 't Hooft-Polyakov monopole was estimated to be of the order of 137 times the mass of W boson. In the simpler SU(2) Yang-Mills-Higgs theory, the mass of W boson is less than 53 GeV, however, the actual mass is found to be $M_w=80.385(15)$GeV experimentally \cite{13}. Therefore, the mass of the 't Hooft-Polyakov monopole is of the order of 11 TeV. \par

The 't Hooft-Polyakov monopole solution found within the SU(2) Yang-Mills-Higgs theory \cite{1}\cite{2} heralded a new era of magnetic monopole research and indeed, different magnetic monopole configurations flourished in the years that follow. Single $n$-monopole \cite{3}, monopole-antimonopole pair (MAP), monopole-antimonopole chain (MAC) and vortex ring solutions \cite{4} are among the most well-known ones. Recently, the existence of monopole configurations with only one-half of the unit topological charge have been reported \cite{5}. In contrast to one-monopoles being point charges, half monopoles are finite line segments located at the origin, extending along the negative $z$-axis. These configurations possess finite total energy even though the electromagnetic gauge potential is singular along the negative $z$-axis. Half monopoles can exist alone or coexist with one-monopoles, forming the so-called one plus half-monopole configuration \cite{6}, which corresponds to the coexistence of a one-monopole of magnetic charge $+1$ at finite $d_z$ along the positive $z$-axis with a half-monopole of magnetic charge $-1/2$ at the origin. \par
In 1977, Y. Nambu predicted the existence of massive string-like configurations within the standard, SU(2)$\times$U(1) Weinberg-Salam model \cite{7}. These configurations are MAP bound by a flux-string of the $Z_0$ field. The total energy of this configuration is finite and the mass of the system is estimated to be in the TeV range. Although the arguments and calculations given weren’t rigorous at the time, this configuration was investigated again in 1984 by Klinkhamer and Manton \cite{8}, and the name "sphaleron" was coined. However, it was believed that due to the quotient space SU(2)$\times$U(1)/U$(1)_{\text{em}}$ allows no non-trivial second homotopy, therefore there exists no topological monopole of interest in the standard Weinberg-Salam model. In 1997, Cho and Maison provided the mathematical proof that magnetic monopole solutions could exist within said model \cite{9} that when the standard Weinberg-Salam model is viewed as a gauged $CP^1$ model, the Higgs field could admit a topologically non-trivial second homotopy, $\pi_2(CP^1)=Z$. This paved the way for constructing realistic magnetic monopole models that are experimentally verifable. Subsequently, numerous monopole configurations, like the monopole-antimonopole chain and vortex ring solutions in this model, were soon established \cite{10}. It is also reported that the standard Weinberg-Salam model could accommodate half-monopole solutions \cite{11}. \par
In this paper, we investigate the one plus half-monopole configuration in the standard Weinberg-Salam model, which is a natural extension of the same configuration found in the SU(2) Yang-Mills-Higgs theory \cite{5}. The equations of motion are solved numerically for all space when the $\phi$-winding number $n$ = 1. The total energy, $E$, and magnetic dipole moment, $\mu_m$, of the solutions are studied by varying the Weinberg angle, $\theta_{\scalebox{.5}{\mbox{W}}}$, from $6.4^\circ$ to $90^\circ$, when the Higgs field self-coupling constant, $\lambda$, is set to 1, and also by varying $\lambda$ from 0 to 36 when $\theta_{\scalebox{.5}{\mbox{W}}}$ is set to $28.74^\circ$. The Higgs field vacuum expectation value, $\zeta$, the unit electric charge, $e$, are both set to unity. The field lines of varies gauge potential and physical fields are constructed and studied as well. 

\section{The Standard Weinberg-Salam Model}
The Lagrangian density in the standard Weinberg-Salam model is given by \cite{9}
	\begin{equation}
	\mathcal{L}=-\left(\mathcal{D}_\mu\pmb\phi\right)^\dagger\left(\mathcal{D}^\mu\pmb\phi\right)-\frac{\lambda}{2}\left(\pmb\phi^\dagger\pmb\phi-\zeta^2\right)^2-\frac{1}{4}\pmb F_{\mu\nu}\cdot\pmb F^{\mu\nu}-\frac{1}{4}G_{\mu\nu}G^{\mu\nu},\label{eqn:L}
	\end{equation}
Here, $\mathcal{D}_\mu$ is the covariant derivative of the SU(2)$\times$U(1) group, which is defined as
	\begin{equation}
	\mathcal{D}_\mu=D_\mu-\frac{i}{2}g'B_\mu=\partial_\mu-\frac{i}{2}g\pmb A_\mu\cdot\pmb\sigma-\frac{i}{2}g'B_\mu,
	\end{equation}
where $D_\mu$ is the covariant derivative of SU(2) group only. In particular, $\pmb\sigma$ is the Pauli vector, whose components are Pauli matrices, which are denoted as $\sigma^a$ down below. The gauge coupling constant, potentials, and electromagnetic fields
of the SU(2) group are given by $g$, $\pmb A_\mu=A^a_\mu\left(\sigma^a/2i\right)$, and $\pmb F_{\mu\nu}=F^a_{\mu\nu}\left(\sigma^a/2i\right)$ respectively, whereas in U(1) group they are $g'$, $B_\mu$, and $G_{\mu\nu}$. The complex scalar Higgs doublet is represented by $\pmb\phi$, the Higgs field mass is $\mu$ and $\zeta=\mu/\sqrt{\lambda}$. The metric used is $-g_{00}=g_{11}=g_{22}=g_{33}=1$. From Eq. (\ref{eqn:L}), the equations of motion are
	\begin{align}
	\mathcal{D}^\mu\mathcal{D}_\mu\pmb\phi&=\lambda\left(\pmb\phi^\dagger\pmb\phi-\zeta^2\right)\pmb\phi,\nonumber\\
	D^\mu\pmb F_{\mu\nu}&=\frac{ig}{2}\left[\pmb\phi^\dagger\pmb\sigma\left(\mathcal{D}_\nu\pmb\phi\right)-\left(\mathcal{D}_\nu\pmb\phi\right)^\dagger\pmb\sigma\pmb\phi\right],\nonumber\\
	\partial^\mu G_{\mu\nu}&=\frac{ig'}{2}\left[\pmb\phi^\dagger\left(\mathcal{D}_\nu\pmb\phi\right)-\left(\mathcal{D}_\nu\pmb\phi\right)^\dagger\pmb\phi\right].\label{eqn:EoM}
	\end{align}
In order to simplify the equations of motion, the Higgs field complex scalar can be expressed as \cite{9}
	\begin{equation}
	\pmb\phi=\abs\Phi\pmb\xi,\quad\pmb\xi^\dagger\pmb\xi=1,\quad\hat\Phi^a=\pmb\xi^\dagger\sigma^a\pmb\xi, 
	\end{equation}
where $\abs\Phi$ is the Higgs modulus, which in the axially symmetrical magnetic ansatz employed in this research is calulated as $g\abs\Phi=\Phi=\sqrt{\Phi_1^2+\Phi_2^2}$. $\pmb\xi$ s a column 2-vector and $\hat\Phi^a$ is the Higgs field unit vector. 

\section{The Axially Symmetric Magnetic Ansatz}
To obtain the one plus half-monopole configuration in standard Weinberg-Salam model, we introduce the the electrically neutral axially symmetric magnetic ansatz \cite{11}, 
	\begin{align}
	gA^a_i=&-\frac{1}{r}\psi_1\hat n^a_\phi\hat\theta_i+\frac{1}{r\sin\theta}P_1\hat n^a_\theta\hat\phi_i+\frac{1}{r}R_1\hat n^a_\phi\hat r_i-\frac{1}{r\sin\theta}P_2\hat n^a_r\hat\phi_i,\nonumber\\
	gA^a_0=&0,\nonumber\\
	g\Phi^a=&\Phi_1\hat n^a_r+\Phi_2\hat n^a_\theta=\Phi\hat\Phi^a,\nonumber\\
	g'B_i=&\frac{1}{r\sin\theta}\mathcal{B}_s\hat\phi_i,\quad g'B_0=0,\nonumber\\
	\pmb\xi=&
		\begin{pmatrix}
		e^{-in\phi}\sin\frac{\alpha}{2}\\
		-\cos\frac{\alpha}{2}
		\end{pmatrix},\hat\Phi^a=\pmb\xi^\dagger\sigma^a\pmb\xi=-\hat h^a,\nonumber\\
	\cos\alpha=&\frac{\Phi_1}{\abs\Phi}\cos\theta-\frac{\Phi_2}{\abs\Phi}\sin\theta,\label{eqn:MA}
	\end{align}
the unit vector, $\hat h^a$, can be expressed as \cite{11} 
	\begin{equation}
	\hat h^a=h_1\hat n^a_r+h_2\hat n^a_\theta=\sin\alpha\cos{n\phi}\delta^{a1}+\sin\alpha\sin{n\phi}\delta^{a2}+\cos\alpha\delta^{a3}, 
	\end{equation}
where, $h_1=\cos(\alpha-\theta)$, $h_2=\sin(\alpha-\theta)$. In magnetic ansatz (\ref{eqn:MA}), $\hat r_i$, $\hat\theta_i$, $\hat\phi_i$ are the unit vectors of ordinary spherical coordinate system, 
	\begin{align}
	\hat r_i&=\sin\theta\cos\phi\,\delta_{i1}+\sin\theta\sin\phi\,\delta_{i2}+\cos\theta\,\delta_{i3},\nonumber\\
	\hat\theta_i&=\cos\theta\cos\phi\,\delta_{i1}+\cos\theta\sin\phi\,\delta_{i2}-\sin\theta\,\delta_{i3},\nonumber\\
	\hat\phi_i&=-\sin\phi\,\delta_{i1}+\cos\phi\,\delta_{i2}, 
	\end{align}
whereas $\hat n^a_r$, $\hat n^a_\theta$, $\hat n^a_\phi$ are the unit vectors for isospin space coordinate,
	\begin{align}
	\hat n^a_r&=\sin\theta\cos n\phi\,\delta^a_1+\sin\theta\sin n\phi\,\delta^a_2+\cos\theta\,\delta^a_3,\nonumber\\
	\hat n^a_\theta&=\cos\theta\cos n\phi\,\delta^a_1+\cos\theta\sin n\phi\,\delta^a_2-\sin\theta\,\delta^a_3,\nonumber\\
	\hat n^a_\phi&=-\sin n\phi\,\delta^a_1+\cos n\phi\,\delta^a_2. 
	\end{align}\par
In the standard Weinberg-Salam model, the physical electromagnetic potential, $\mathcal{A}_\mu$,and neutral field, $\mathcal{Z}_\mu$, are related to the gauge fields through \cite{12} 
	\begin{equation}
		\begin{bmatrix}
		\mathcal{A}_\mu\\
		\mathcal{Z}_\mu
		\end{bmatrix}=
		\begin{bmatrix}
		\cos\theta_W&\sin\theta_W\\
		-\sin\theta_W&\cos\theta_W
		\end{bmatrix}
		\begin{bmatrix}
		B_\mu\\
		A'^3_\mu
		\end{bmatrix}=\frac{1}{\sqrt{g^2+g'^2}}
		\begin{bmatrix}
		g&g'\\
		-g'&g
		\end{bmatrix}
		\begin{bmatrix}
		B_\mu\\
		A'^3_\mu
		\end{bmatrix},
	\end{equation}
where, $\theta_{\scalebox{.5}{\mbox{W}}}=\cos^{-1}\left(\frac{g}{\sqrt{g^2+g'^2}}\right)$. Then, it is obvious that 
	\begin{align}
	e\mathcal{A}_\mu&=\left(\cos^2\theta_{\scalebox{.5}{\mbox{W}}}g'B_\mu+\sin^2\theta_{\scalebox{.5}{\mbox{W}}}gA^{'3}_\mu\right),\nonumber\\
	e\mathcal{Z}_\mu&=-\cos\theta_{\scalebox{.5}{\mbox{W}}}\sin\theta_{\scalebox{.5}{\mbox{W}}}\left(-g'B_\mu+gA^{'3}_\mu\right),
	\end{align}
where $e=\frac{gg'}{\sqrt{g^2+g'^2}}$ is the unit electric charge \cite{12}. Furthermore, $\theta_{\scalebox{.5}{\mbox{W}}}$ can be expressed through the relation $M_{\scalebox{.5}{\mbox{W}}}/M_{\scalebox{.5}{\mbox{Z}}}=\cos\theta_{\scalebox{.5}{\mbox{W}}}$, where $M_{\scalebox{.5}{\mbox{W}}}$ and $M_{\scalebox{.5}{\mbox{Z}}}$ are the masses of the W and Z bosons. With the experimental values for those, where $M_{\scalebox{.5}{\mbox{W}}}=80.385(15)$GeV and $M_{\scalebox{.5}{\mbox{Z}}}=91.1876(21)$GeV \cite{13}, $\theta_{\scalebox{.5}{\mbox{W}}}$ can be calculated to be $28.74^\circ$. \par
To investigate the magnetic property of the solution, the gauge transformation, $U$, was chosen such that $U\Phi^a=\Phi'^a=\delta^3_a$. Specifically, 
	\begin{equation}
	U=-i
		\begin{bmatrix}
		\cos\frac{\alpha}{2}&\sin\frac{\alpha}{2}e^{-in\phi}\\
		\sin\frac{\alpha}{2}e^{in\phi}&-\cos\frac{\alpha}{2}
		\end{bmatrix}=\cos\frac{\Theta}{2}+i\hat u^a_r\sigma^a\sin\frac{\Theta}{2},
	\end{equation}
where $\Theta=-\pi$ and $\hat u^a_r=\sin\frac{\alpha}{2}\cos n\phi\,\delta^a_1+\sin\frac{\alpha}{2}\sin n\phi\,\delta^a_2+\cos\frac{\alpha}{2}\,\delta^a_3$. Upon applying the transformation, the third component of the transformed gauge field is then
	\begin{align}
	gA^{'a}_\mu&=-gA^a_\mu-\frac{2}{r\sin\theta}\left[P_1\sin\left(\theta-\frac{\alpha}{2}\right)+P_2\cos\left(\theta-\frac{\alpha}{2}\right)\right]\hat u^a_r\hat\phi_\mu-\partial_\mu\alpha\hat u^a_\phi-\frac{2n\sin\frac{\alpha}{2}}{r\sin\theta}\hat u^a_\theta\phi_\mu\nonumber\\
	\Rightarrow gA^{'3}_\mu&=\frac{1}{r\sin\theta}\left[P_1h_2-P_2h_1-n\left(1-\cos\alpha\right)\right]\hat\phi_\mu=\frac{A_1}{r\sin\theta}\hat\phi_\mu,
	\end{align}
here, the $gA'^3_\mu$ produced is precisely the negative gauge potential of the 't Hooft electromagnetic tensor \cite{10}, $\hat F_{\mu\nu}=\hat\Phi^aF^a_{\mu\nu}-\frac{1}{g}\varepsilon^{abc}\hat\Phi^aD_\mu\hat\Phi^bD_\nu\hat\Phi^c$ \cite{1}. For this reason, the U(1) and SU(2) magnetic field could be expressed as
	\begin{align}
	g'B^{\text{U(1)}}_i&=-\frac{g'}{2}\varepsilon^{ijk}G_{jk}=-\varepsilon^{ijk}\partial_j\left\{nB_1\sin\theta\right\}\partial_k\phi,\\
	gB^{\text{SU(2)}}_i&=-\frac{g}{2}\varepsilon^{ijk}\hat F_{jk}=\varepsilon^{ijk}\partial_j\left(gA'^3_k\right)=-\varepsilon^{ijk}\partial_j\left\{A_1\sin\theta\right\}\partial_k\phi,
	\end{align}
respectively. The magnetic field lines of respective gauge fields could then be constructed by drawing the contour lines of the terms in curly brackets. \par
The energy density of the system is obtained from the energy-momentum tensor, $T^{\mu\nu}$, and in the electrically neutral monopole configuration, it has the form \cite{11}
	\begin{equation}
	e^2\varepsilon_n=T^{00}=\cos^2\theta_{\scalebox{.5}{\mbox{W}}}\varepsilon_0+\sin^2\theta_{\scalebox{.5}{\mbox{W}}}\varepsilon_1+\varepsilon_H. 
	\end{equation}
In particular, the three components $\varepsilon_0$, $\varepsilon_1$ and $\varepsilon_H$ are the corresponding energy density for U(1) gauge field, SU(2) gauge field and Higgs field respectively, which are given by 
	\begin{align}
	\varepsilon_0=&\frac{g'^2}{4}G_{ij}G_{ij},\varepsilon_1=\frac{g^2}{4}F^a_{ij}F^a_{ij},\nonumber\\
	\varepsilon_H=&\sin^2\theta_{\scalebox{.5}{\mbox{W}}}\partial^i\Phi\partial_i\Phi+\sin^2\theta_{\scalebox{.5}{\mbox{W}}}\Phi^2(\mathcal{D}^i\pmb\xi)^\dagger(\mathcal{D}_i\pmb\xi)+\frac{\lambda}{2}(\sin^2\theta_{\scalebox{.5}{\mbox{W}}}\Phi^2-\zeta^2)^2,\nonumber\\
	\left(\mathcal{D}^i\pmb\xi\right)^\dagger\left(\mathcal{D}_i\pmb\xi\right)=&\frac{1}{4}\partial^i\alpha\partial_i\alpha+\frac{n^2\left(1-\cos\alpha\right)}{2r^2\sin^2\theta}+\frac{n}{2}\left(1-\cos\alpha\right)\left(g'B^i\right)\partial_i\phi\nonumber\\
	&+\frac{1}{2}\left[\hat n^a_\phi\partial^i\alpha+n\partial^i\phi\left(\hat n^r_a\cos\theta-\hat n^a_\theta\sin\theta-\hat h^a\right)\right]\left(gA^a_i\right)\nonumber\\
	&+\frac{1}{4}\left(gA^{ai}\right)\left(gA^a_i\right)-\frac{1}{2}\left(g'B^i\right)\left(gA^a_i\right)\hat h^a+\frac{1}{4}\left(g'B^i\right)\left(g'B_i\right).
	\end{align}
The energy density itself contains singularities along the negative $z$-axis due to the presence of the half-monopole, however, it is in fact integrable and therefore the total energy of the system is finite. For this reason, we introduced the weighted energy density defined as $\varepsilon_{\scalebox{.5}{\mbox{W}}}=r^2\sin\theta\varepsilon_n$. The total energy, $E$, could then be calculated from the following integral 
	\begin{equation}
	E=\frac{e}{4\pi}\int\varepsilon_n d^3x=\frac{e}{4\pi}\int\int\int\varepsilon_nr^2\sin\theta drd\theta d\phi=\frac{e}{2}\int\int\varepsilon_{\scalebox{.5}{\mbox{W}}}drd\theta.
	\end{equation}\par
The electromagnetic dipole moment, $\mu_m$, can be calculated by using the boundary conditon of the electromagnetic gauge potential at large $r$ \cite{11},
	\begin{equation}
	\mathcal{A}_\mu\rightarrow\frac{1}{e}\left(g'B_i\right)=\frac{1}{e}\mathcal{B}_s\partial_i\phi=-\frac{\hat\phi_i}{r\sin\theta}\left(\frac{\mu_m\sin^2\theta}{r}\right).
	\end{equation}
Hence, $r\mathcal{B}_s=-e\mu_m\sin^2\theta$ and by plotting the numerical result for $r\mathcal{B}_s$, the magnetic dipole moment can be read off in unit of $1/e$ at $\theta=\pi/2$. 

\section{Numerical Procedure}
In magnetic ansatz (\ref{eqn:MA}), all profile functions $\psi_1$, $P_1$, $R_1$, $P_2$, $\Phi_1$, $\Phi_2$, and $\mathcal{B}_s$ introduced are functions of $r$ and $\theta$. The magnetic ansatz (\ref{eqn:MA}) is substituted into the equations of motion from Eqs. (\ref{eqn:EoM}) and the set of equations are then reduced to 7 coupled non-linear second order partial differential equations. These coupled equations are then solved using Maple and MATLAB by fixing boundary conditions at small distances ($r\rightarrow0$), large distances ($r\rightarrow\infty$), and along the positive and negative $z$-axis when $\theta=0$ and $\pi$. \par
The boundary conditions used at small $r$ is the vacuum trivial solution
	\begin{align}
	\psi_1(0,\theta)=P_1(0,\theta)=R_1(0,\theta)=P_2(0,\theta)=\mathcal{B}_s(0,\theta)=&0,\nonumber\\
	\sin\theta\Phi_1(0,\theta)+\cos\theta\Phi_2(0,\theta)=&0,\nonumber\\
	\partial_r[\cos\theta\Phi_1(r,\theta)-\sin\theta\Phi_2(r,\theta)]|_{r=0}=&0,
	\end{align}
and asymptotically when $r\rightarrow\infty$, we have the self-dual solution 
	\begin{align}
	\psi_1(\infty,\theta)&=3/2, R_1(\infty,\theta)=0, \nonumber\\
	P_1(\infty,\theta)&=\sin\theta-\frac{1}{2}\sin\frac{\theta}{2}(1+\cos\theta), \nonumber\\
	P_2(\infty,\theta)&=\cos\theta-\frac{1}{2}\cos\frac{\theta}{2}(1+\cos\theta), \nonumber\\
	\Phi_1(\infty,\theta)&=\zeta\cos\frac{\theta}{2}, \Phi_2(\infty,\theta)=\zeta\sin\frac{\theta}{2}, \nonumber\\
	\mathcal{B}_s(\infty,\theta)&=-\frac{1}{2}(1-\cos\theta). 
	\end{align}
Similarly, along the $z$-axis for $\theta=0$ and $\pi$,
	\begin{align}
	\partial_\theta\psi_1(r,\theta)|_{\theta=0}&=R_1(r,0)=P_1(r,0)=P_2(r,0)=\partial_\theta\Phi_1(r,\theta)|_{\theta=0}=\Phi_2(r,0)=\mathcal{B}_s(r,0)=0,\nonumber\\
	\partial_\theta\psi_1(r,\theta)|_{\theta=\pi}&=R_1(r,\pi)=P_1(r,\pi)=\partial_\theta P_2(r,\theta)|_{\theta=\pi}=\Phi_1(r,\pi)=\partial_\theta\Phi_2(r,\theta)|_{\theta=\pi}=\partial_\theta\mathcal{B}_s(r,\theta)|_{\theta=\pi}=0.
	\end{align}\par
The seven coupled second-order partial differential equations were then converted into a system of nonlinear equations using finite difference approximation method, which were then discretized onto a non-equidistant grid of 70$\times$60 covering the integration regions $0\leq\bar{x}\leq1$ and $0\leq\theta\leq\pi$, where $\bar{x}$ is the compactified coordinate $\bar{x}=\frac{r}{r+1}$, converting the radial $r$-axis from $0\rightarrow\infty$ to $0\rightarrow1$. \par
The following substitutions were made to all the partial derivatives in the equations
	\begin{equation}
	\partial_r\rightarrow\left(1-\bar{x}\right)^2\partial_{\bar{x}},\quad\frac{\partial^2}{\partial r^2}\rightarrow\left(1-\bar{x}\right)^4\frac{\partial^2}{\partial\bar{x}^2}-2\left(1-\bar{x}\right)^3\frac{\partial}{\partial\bar{x}},
	\end{equation}
in order to construct the Jacobian sparsity pattern of the system using Maple, which was used to optimize the numerical calculations. Finally, the system of nonlinear equations is then solved numerically by MATLAB using the said boundary conditions,Jacobian sparsity pattern constructed, the trust-region-reflective algorithm, and a good initial starting solution. 

\section{Results and Discussion}
	\begin{figure}[h]
	\includegraphics[width=\linewidth]{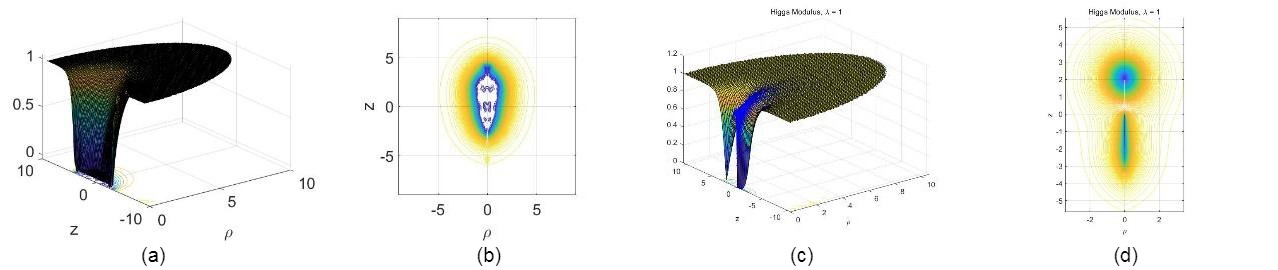}
	\caption{(a) 3D plot and (b) contour line plot of Higgs modulus for one plus half-monopole in the standard Weinberg-Salam model when $\theta_{\scalebox{.5}{\mbox{W}}}=90^\circ$, $\lambda=1$, (a) 3D plot and (b) contour line plot of Higgs modulus for one plus half-monopole in SU(2) Yang-Mills-Higgs theory when $\lambda=1$.}
	\end{figure}
In Fig. 1, the 3D Higgs modulus and its contour plots for one plus half-monopole configuration in both standard Weinberg-Salam model and SU(2) Yang-Mills-Higgs theory are shown. In SU(2) Yang-Mills-Higgs theory, the half-monopole is located at the origin and extending itself along the negative $z$-axis, whereas the one-monopole is located somewhere along the positive $z$-axis. They are two distinct entities with a clear, measureable pole separation, $d_z$, as shown in Fig. 1(c) and (d). However, in the case of standard Weinberg-Salam model, while the pole separation becomes larger, the poles are now connected by a flux tube, forming a bound state, as in Fig. 1(a) and (b). The pole separation can no longer be accurately measured. The flux tube connecting both gets boarder as it extends itself from the half-monopole towards the one-monopole. \par
	\begin{figure}[h]
	\includegraphics[width=\linewidth]{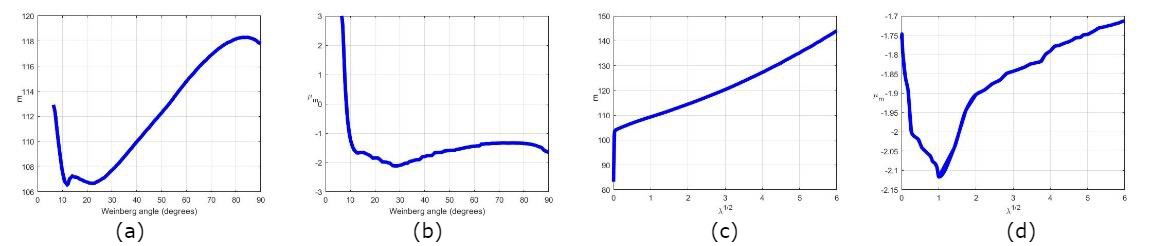}
	\caption{Plots of one plus half-monopole configuration for (a) $E$ versus $\theta_{\scalebox{.5}{\mbox{W}}}$, (b) $\mu_m$ versus $\theta_{\scalebox{.5}{\mbox{W}}}$, (c) $E$ versus $\lambda$, (d) $\mu_m$ versus $\lambda$.}
	\end{figure}
Figure 2(a) and (b) shows plots of $E$ and $\mu_m$ of one plus half-monopole configuration versus $\theta_{\scalebox{.5}{\mbox{W}}}$, $6.4^\circ\leq\theta_{\scalebox{.5}{\mbox{W}}}\leq90^\circ$, at $\lambda=1$. From Fig. 2 (a), $E$ descends from $E=112.8967$ until $\theta_{\scalebox{.5}{\mbox{W}}}=12^\circ$ where it reaches the first local minimum with $E=106.5568$, then rises and at $\theta_{\scalebox{.5}{\mbox{W}}}=14^\circ$ with $E=107.2574$ where it reaches a local maximum. The second local minimum appears at $\theta_{\scalebox{.5}{\mbox{W}}}=23^\circ$ with $E=106.6907$. The energy keeps increasing logarithmically until $\theta_{\scalebox{.5}{\mbox{W}}}=85^\circ$ with $E=118.2912$ and decreases to $E=117.8027$ when $\theta_{\scalebox{.5}{\mbox{W}}}=90^\circ$. In Fig. 2(b), $\mu_m$ of the system is initially positive and starts to decrease with increasing $\theta_{\scalebox{.5}{\mbox{W}}}$ and reaches 0 when $\theta_{\scalebox{.5}{\mbox{W}}}=8.37^\circ$, then it's negative for the remainder of the plot. Two local extrema occur at $\theta_{\scalebox{.5}{\mbox{W}}}=23^\circ$ and $74^\circ$, with $\mu_m=-2.1135$ and $-1.3335$. Fig. 2(c) and (d) shows plots of $E$ and $\mu_m$ versus $\lambda$, $0\leq\lambda\leq36$, at $\theta_{\scalebox{.5}{\mbox{W}}}=28.74^\circ$. From Fig. 2(c) $E$ undergoes a sharp increase from $\lambda=0$ to $0.03$ before increases monotonically with increasing $\lambda$. In Fig. 2(d), $\mu_m$ decreases monotonically between $\lambda=0$ to 1 with $\mu_m=-2.1168$ being the minimum value, then increases with increasing $\lambda$. \par
	\begin{figure}[h]
	\includegraphics[width=\linewidth]{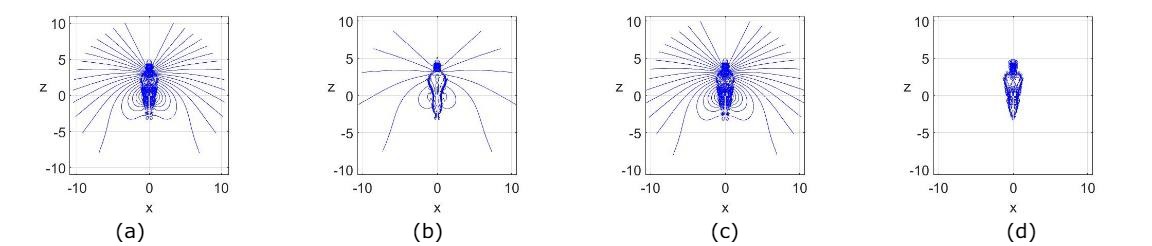}
	\caption{Plots of field lines of (a) EM field, (b) U(1) gauge field, (c) SU(2) gauge field and (d) neutral field.}
	\end{figure}
Figure 3 shows the field lines of various gauge fields presented in this configuration. From Fig. 3(a), the field lines of the electromagnetic field clearly indicate that in this configuration, the positive charge is distributed among the one-monopole and the upper half of the flux tube, the lower half of the flux tube and the half-monopole at the origin is negatively charged. In Fig. 3(b) and (c), similar observations can be made, however, in the case of U(1) gauge field, the intensity is much lower and there is no indication that electromagnetic current loops are present. This is in contrast to the work done in Ref. \cite{10}, where the monopole-antimonopole pair configuration is investigated, in which case, the configuration does not present in the U(1) gauge field at all and at the same time, an electromagnetic current loop is reported. The shape of electromagnetic field lines in Fig. 3(a) shows striking resemblence to the SU(2) magnetic field lines in Fig. 3(c). This indicates the SU(2) contribution significantly outweighs the one from U(1) gauge field. In Fig. 3(d), the neutral field appears to be confining itself within the flux tube as there’s no field line going out, which is similar to the case of monopole-antimonopole pair configuration \cite{7}\cite{10}. 

\section{Conclusion}
In this work, we have investigated the one plus half-monopole configuration of the standard Weinberg-Salam model, which is a natural extension from the same configuration in the SU(2) Yang-Mills-Higgs theory reported previously. Similar to the monopole-antimonopole pair in the standard Weinberg-Salam model, this configuration is also bounded by a neutral flux string, forming a tube-like structure connecting both the monopole and half-monopole. However, there is no electromagnetic current loop circulating the pair of one-monopole and half-monopole in the U(1) gauge field. The configuration is present in the U(1) gauge field, however, the SU(2) contribution significantly outweighs the former. Mathematically, even if there are innate singularities presented in the system along the negative $z$-axis due to the presence of the half-monopole, the system possesses finite total energy, $E$, and magnetic dipole moment, $\mu_m$. However, they possess behaviors that are drastically different from their counterpart in SU(2) Yang-Mills-Higgs theory. More detailed work and further study of this configuration with higher $\phi$-winding number, $n$, broader range of the Higgs self-coupling constant, $\lambda$, and electric charge introduced will be reported in a separate paper. 

\section{Acknowledgements}
The authors would like to thank School of Physics USM, Bridging Fund (Grant No: 304/PFIZIK/6316278) and the organizing committee of 14th APPC, for funding the research and conference expenses.

\end{document}